\documentclass[notitlepage,aps,prx,reprint,twocolumn,longbibliography,superscriptaddress]
{revtex4-1}
\usepackage{graphicx}
\usepackage{amsmath}
\usepackage{amssymb}
\usepackage{comment}
\usepackage[colorlinks, allcolors=blue]{hyperref}
\usepackage[all]{hypcap}
\usepackage[mathlines]{lineno}
\usepackage{physics}
\usepackage{wrapfig}
\usepackage{lipsum}
\usepackage[normalem]{ulem}
\usepackage{siunitx}
\usepackage{bbold}
\usepackage{animate}

\newcommand{\pref}[2]{\hyperref[#1]{\ref{#1}(#2)}}
\newcommand{\preff}[2]{\hyperref[#1]{\ref{#1}#2}}
\newcommand{\eqpref}[1]{\hyperref[#1]{(\ref{#1})}}

\newcommand{\squig}{{\raise.17ex\hbox{$\scriptstyle\sim$}}}

\begin{document}
	\title{Interaction-assisted topological pumping in few- and many-atom Rydberg arrays}
\author{Chenxi Huang}
\affiliation{Department of Physics, University of Illinois at Urbana-Champaign, Urbana, IL 61801-3080, USA}
\affiliation{Department of Physics, The Pennsylvania State University, University Park, Pennsylvania 16802, USA}
	\author{Tao Chen}
        \thanks{Present address: School of Physics, Xi'an Jiaotong University, Xi'an, China}
	\affiliation{Department of Physics, University of Illinois at Urbana-Champaign, Urbana, IL 61801-3080, USA}
    \affiliation{Department of Physics, The Pennsylvania State University, University Park, Pennsylvania 16802, USA}
\author{Qian Liang}
\author{Matthew A.~Krebs}
\author{Ethan Springhorn}
\author{Ruiyu Li}
\author{Mingsheng Tian}
\affiliation{Department of Physics, The Pennsylvania State University, University Park, Pennsylvania 16802, USA}
 \author{Kaden R.~A.~Hazzard}
        \affiliation{Department of Physics and Astronomy, Rice University, Houston, TX 77005, USA}
        \affiliation{Smalley-Curl Institute, Rice University, Houston, TX 77005, USA}
	\author{Jacob P. Covey}
	\affiliation{Department of Physics, University of Illinois at Urbana-Champaign, Urbana, IL 61801-3080, USA}
	\author{Bryce Gadway}
\email{bgadway@psu.edu}
 \affiliation{Department of Physics, The Pennsylvania State University, University Park, Pennsylvania 16802, USA}
	\affiliation{Department of Physics, University of Illinois at Urbana-Champaign, Urbana, IL 61801-3080, USA}
	\date{\today}
 
\begin{abstract}
Topology can imbue lattice systems with special properties, notably the presence of robust eigenstates living at their boundary. Through dimensional reduction, the robust bulk band topology of, e.g., the integer quantum Hall system can be mapped onto similarly robust charge-pumping dynamics of a topological pump living in one lower dimension. Recent studies have uncovered a rich influence of interactions on the dynamics of topological pumps in nonlinear systems, including the robust pumping of self-bound solitons. These striking observations in classical nonlinear photonics have raised a number of questions, chiefly if and how this phenomenology persists in strongly correlated quantum systems and in the few-body limit. Here, using few- and many-atom arrays, we explore how dipolar interactions impact the dynamics of topological population pumping along a Rydberg synthetic dimension.
In the few-body limit, we find that dipolar interactions lead to self-bound states that are efficiently pumped along the synthetic dimension, described by an emergent pair-state topological pump. We find that this interaction-assisted pumping persists in many-atom arrays, with a sharpened dependence on the dipolar interaction strength that stems from the enhanced spatial connectivity. These Rydberg-based studies on interaction-assisted topological pumping help connect observations from classical nonlinear photonics to the few-body quantum limit and pave the way for studies of new strongly correlated quantum pumping phenomena.
\end{abstract}

\maketitle

\section{Introduction}

Topological pumping~\cite{citroThoulessPumpingTopology2023} refers to the quantized transport of charge via adiabatic, cyclic modulations over control parameters, without the requirement for an external bias. This concept, originally proposed by Thouless~\cite{Thouless1983}, relies on the topology of the energy band to determine the amount of charge transported per cycle, i.e., the Chern number, making the process robust against perturbations.  The ability to transport charge in the absence of the external potential bias also suggests a promising alternative current standard, capable of reducing energy dissipation~\cite{Pekola2013}. Mathematically, Thouless pumping can be viewed as a one-dimensional reduction of the two-dimensional quantum Hall effect by mapping one of the momentum coordinates to time. Experimentally, such quantized transport phenomena have been observed in various platforms, including ultracold atoms~\cite{Lohse2016,Nakajima2016,Lu2016}, photonics~\cite{Kraus2012,Cerjan2020}, superconducting circuits~\cite{Tao2025emulating,Deng2024,Liu2025}, and more~\cite{Grinberg2020,Xia2021}. To clarify the mechanisms for topological protection in the presence of added perturbations, experiments have begun to explore topological pumping in the presence of disorder~\cite{Cerjan2020,Liu2025} and quasiperiodic modulations~\cite{nakajimaCompetitionInterplayTopology2021} as well as under added dissipation~\cite{Fedorova2020,Ravets2025}.

\begin{figure}[ht]
    \includegraphics[width=0.5\textwidth]{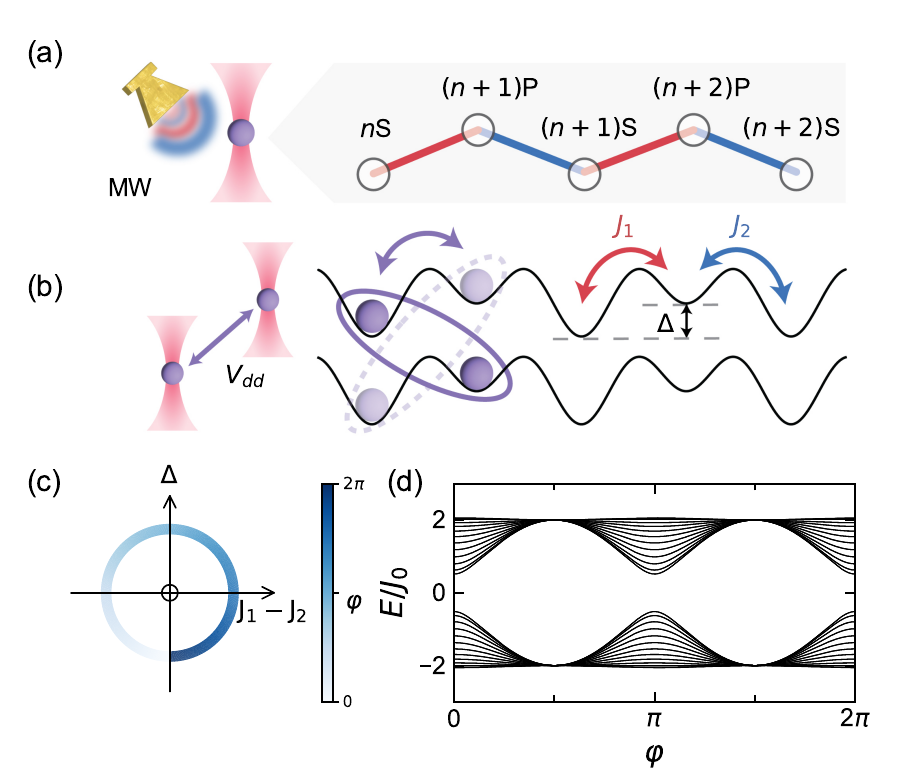}
\caption{\textbf{Topological pumping along a Rydberg synthetic lattice with dipolar interactions.} 
    \textbf{(a)}~Multiple microwave tones are applied to Rydberg atoms (left) to create an effective tight-binding model (right) of resonantly-coupled internal states.
    \textbf{(b)}~Depiction of the effective single-particle Rice-Mele model (Eq.~\ref{eqn:sRM}) with staggered intra- ($J_1$) and inter-cell ($J_2$) hopping and potential landscape modulation $\Delta$. For atoms in nearby tweezer traps, dipolar interactions ($V_{\rm{dd}}$) lead to the correlated anti-hopping of Rydberg electrons along the synthetic dimension. The Rice-Mele lattice parameters are controlled as $J_{1/2}(\varphi) = J_0 (1\pm\cos(\varphi))$ and $\Delta(\varphi) =\Delta_0\sin(\varphi)$, where $\varphi$ is the pump phase.
    \textbf{(c)}~Pumping trajectory as depicted in the $(J_1 - J_2)$ vs. $\Delta$ parameter space, enclosing the gap-closing condition at the origin.
    \textbf{(d)}~The instantaneous one-atom energy band structure plotted against the pump phase $\varphi$, shown for $\Delta_0 = 2 J_0$.
    }
\label{FIG:fig1}
\end{figure}

Interparticle interactions are also important perturbations to topological pumps, establishing an additional energy scale and opening up the possibility of new pumping phenomena in correlated systems. Recent efforts have explored how interactions modify the behavior of topological pumps, from the stable pumping of self-bound solitons in the context of mean-field or classical nonlinearities~\cite{Jurgensen2021, Jurgensen2023,Mostaan2022,Fu2022,Cao2025,JWP-pumps} to the influence of strong Hubbard interactions on topological pumping in ultracold atom experiments~\cite{FHM-2018,Walter2023,Viebahn2024}, including the possibility of few- and many-body pumps enabled by interactions~\cite{CHLee-2020,CHLee-Tilt,CHLee-BIC,Kuno2020,FHM-2022-splitting,liuCorrelatedTopologicalPumping2023,Viebahn2024,lam2024interactioninducedsplittingdirac}. More broadly, exploring interacting topological pumps promises to enable robust quantum state transformations in many-body systems~\cite{Zhu-Gates}, to enable technological applications related to frequency conversion, quantum state preparation, and sensing~\cite{Martin-Top-Freq-Convert,Long-Boosting}, and to uncover new classes of many-body topological phenomena~\cite{Zhu-reversal,Jurg-multi}.

The local, time-dependent control required for topological pumping can be achieved naturally in experiments based on synthetic dimensions~\cite{celi2014synthetic,Ozawa2019,Hazz-SynthDim-Rev,Arguello-Luengo2024,Fabre_2024,Yuan-2025}, where tunable tight-binding models are engineered by controlling, e.g., transitions between the discrete states of atoms. While Rydberg atoms in synthetic dimensions have recently been used to explore topology~\cite{Kanungo2022,Lu2024} and quantized charge pumps~\cite{Trautmann2024}, these studies have been restricted to the case of individual, non-interacting particles. Here, we use few- and many-site arrays of Rydberg atoms to explore how strong interactions influence the topological pumping of Rydberg electrons along an internal state synthetic dimension with five sites. While the dipolar exchange interactions of Rydberg atoms, equivalent to pairwise anti-correlated hopping along the synthetic dimension~\cite{Sundar2018,Chen2024,Chen2024a,Chen2025}, are microscopically distinct from the on-site nonlinearities and interactions of photonic~\cite{Jurgensen2021} and cold atom~\cite{Preiss-QWalk,Walter2023} experiments, we observe an analogous emergence of self-bound states that are topologically pumped in a collective fashion. In the case of just two atoms, we connect our observations to the emergence of an effective pair-state Thouless pump with reduced positional spreading.
Extending to few- and many-atom arrays, we observe a pumping behavior that depends on the collective strength of the dipolar interactions, connecting to the behavior of topologically pumped solitons in nonlinear photonics~\cite{Jurgensen2021,Mostaan2022,Jurg-2022}. These studies pave the way for future studies on interaction-enabled pumping phenomena in Rydberg synthetic dimensions.

\section{Experimental implementations}

\subsection{Rydberg synthetic dimensions}

As described in our previous studies~\cite{Chen2024,Chen2024a,Chen2025}, we encode Rydberg states as lattice sites, with microwave electric fields applied to drive transitions between these states. Global microwave drives are used to couple lattice sites, where the strengths of the applied microwaves directly control the effective tunneling rates. The detunings of the microwave tones from the individual state-to-state transitions are used to engineer the effective potential energy landscape of the synthetic lattice. Each state-to-state transition is characterized by a distinct resonance and is spectroscopically well-isolated, allowing for the precise control of individual tunneling links.

We use a set of Rydberg states as shown in Fig.~\pref{FIG:fig1}{a} to encode a few-site Rice-Mele (RM) Hamiltonian
\begin{eqnarray}
    H(\varphi) = & - &\sum_m \left[J_1 (\varphi)b^{\dagger}_m a^{\phantom\dagger}_m +J_2(\varphi)a^{\dagger}_{m+1} b^{\phantom\dagger}_m + {\rm h.c.}\right] \nonumber\\
    & + &\frac{\Delta(\varphi)}{2}\sum_m(a^{\dagger}_ma^{\phantom\dagger}_m - b^{\dagger}_mb^{\phantom\dagger}_m)
\label{eqn:sRM}
\end{eqnarray}
where $a^{\dagger}_m$ ($a^{\phantom\dagger}_m$) and $b^{\dagger}_m$ ($b^{\phantom\dagger}_m$) are the creation (annihilation) operators for the odd and even sites of the $m$-th unit cell. $J_1(\varphi)$ and $J_2(\varphi)$ respectively denote the intra- and inter-cell tunnelings, while $\Delta(\varphi)$ is the energy difference between the two sites within one unit cell, depicted in Fig.~\pref{FIG:fig1}{b}. $\varphi$ is the modulation phase of the RM model, with a time-dependence $\varphi (t) = 2\pi\omega t + \varphi_0$ (pump frequency $\omega$, initial phase $\varphi_0$) leading to the cyclic, periodic modulation of the RM model parameters (Fig.~\pref{FIG:fig1}{c}) and energy spectrum (Fig.~\pref{FIG:fig1}{d}) of our pump. To note, the same pump trajectory is utilized for all single-atom, few-atom, and many-atom studies. More details on the pump trajectory and its experimental implementation can be found in Appendix~\ref{appendixA}.

Our synthetic lattice studies involving few-atom arrays (singles and dimers, or pairs) and many-atom arrays (singles, dimers, trimers, and two-dimensional arrays), are separately realized in experimental systems based on $^{39}$K Rydberg atom arrays and $^{87}$Rb Rydberg atom arrays, respectively. These distinct experimental systems utilize distinct sets of Rydberg levels for the implementation of the RM models, which we now explicitly detail.

In our experiments based on $^{39}$K~\cite{JacksonPRR,Chen2024}, we perform one- and two-atom synthetic lattice studies by probabilistically loading a one-dimensional dimerized configuration of tweezer traps.
Single atoms and dimers (pairs) are identified via post-selection based on fluorescence images taken prior to the science portion of the experiments. The trapped, cooled, and optically pumped $^{39}$K atoms are globally excited to the Rydberg state $\ket{0}\equiv\ket{42S_{1/2},m_J=1/2}$. A five-state synthetic lattice is constructed through the application of multiple microwave tones in the range of 44 to 52 GHz, realizing an effective tight-binding lattice structure described by Eq.~\ref{eqn:sRM} and depicted in Figs.~\pref{FIG:fig1}{a,b}. The lattice is composed of the states $\ket{41S_{1/2}}$, $\ket{41P_{3/2}}$, $\ket{42S_{1/2}}$, $\ket{42P_{3/2}}$, and $\ket{43S_{1/2}}$, with different $m_J$ sublevels used to access different signs of the interaction $V$.

Our synthetic lattice studies on many-atom systems are implemented in two-dimensional tweezer arrays of individual $^{87}$Rb atoms. The one-, two-, and three-atom configurations of $^{87}$Rb are also identified through post-selection, while our many-atom triangular arrays are probabilistically loaded as discussed later on. For the $^{87}$Rb studies, our initialized state is $\ket{62S_{1/2},m_J=1/2}$, and a five-site synthetic lattice (involving the states $\ket{61S_{1/2}}$, $\ket{61P_{3/2}}$, $\ket{62S_{1/2}}$, $\ket{62P_{3/2}}$, and $\ket{63S_{1/2}}$, all with $m_J = 1/2$) is engineered by the application of microwave frequency tones in the range of 15.5 to 16.7~GHz.

\begin{figure}[t]
	\includegraphics[width=0.5\textwidth]{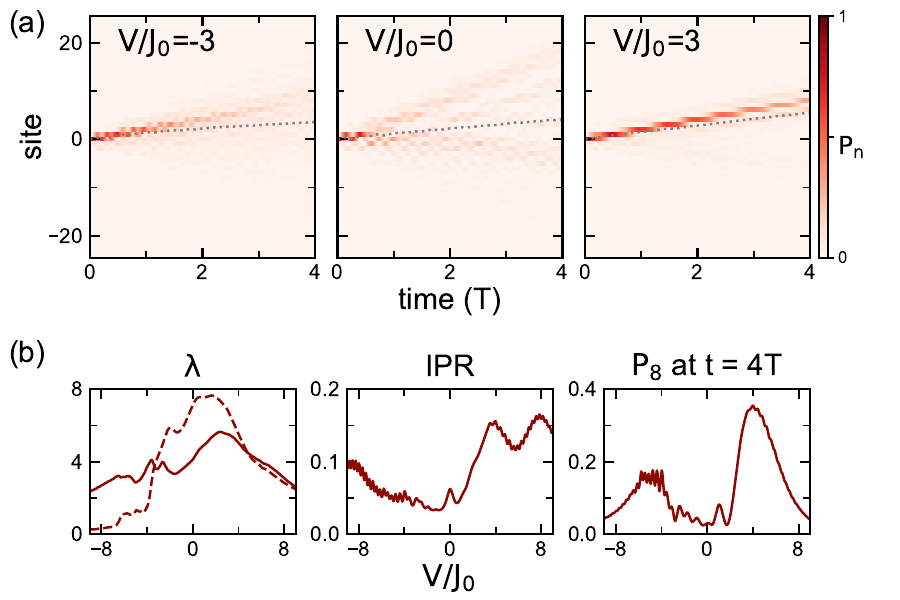}
	\caption{\textbf{Influence of dipolar interactions on two-body pumping dynamics.} 
        \textbf{(a)}~Pumping dynamics are simulated with 50-site systems and initiated at the center. Plots from left to right correspond to interaction-to-hopping ratios of $V/J_0 = -3, 0, 3$. Gray dashed lines indicate the center of mass position over time, and the corresponding line should have a slope of 2 for an ideal pump. All three cases show directional transport, but population spreads across the system faster for $V/J_0 = 0$.
        \textbf{(b)}~Interaction dependence of the simulated pumping behavior of an atom pair, characterized at a fixed time $t = 4T$. Left: The tendency to pump is captured by the center of mass position $\lambda$, plotted vs. $V/J_0$. The solid line is for preparing both atoms at site $0$, while the dashed line assumes perfect preparation of the single-particle lower-band Wannier state. Center: The tendency to stay localized, characterized by the inverse partition ratio (IPR).
        The tendency to both pump and remain localized can be captured by the population $2n$ sites ($n$ unit cells) away from the initial site after $n$ pumping periods. Right: The population at site 8, $P_8$, plotted vs. $V/J_0$.
        }      
\label{FIG:fig2}
\end{figure}

The topological pumping explored in this study is described by the trajectory in Fig.~\pref{FIG:fig1}{b} with cyclic modulations $J_{1/2}(\varphi) = J_0 (1\pm\sin(\varphi))$ for tunneling between sites and $\Delta(\varphi) =\Delta_0\cos(\varphi)$ for the potential energy difference.
For the simulations of Fig.~\ref{FIG:fig2} and for the experimental data with $^{39}$K, we set the average tunneling rate $J_0/h = 0.75(1)$~MHz and maximum energy difference $\Delta_0/h = 2J_0/h = 1.5(2)$~MHz. When experimentally exploring the interaction dependence with $^{39}$K, we vary the ratio $V/J_0$ by directly changing the atomic interaction strength by changing the spacing between atoms.
In this case, the pump modulation phase is varied in time as $\varphi (t) = 2\pi \omega t + \varphi_0$ with period $T = 1.33~\mu$s, modulation frequency $\omega = 2\pi/T = J_0$ and initial phase $\varphi_0 = \pi$; a theory analysis of adiabaticity can be found in Appendix~\ref{appendixD} and Fig.~\ref{FIG:figS2}. Ideally, longer modulation periods ensure more ideal pumping, but we settle for approximate adiabaticity due to a limited experimental time window.

For the studies based on $^{87}$Rb arrays, the modulation parameters are $\Delta_0 = 2.8J_0$ and $\omega = 2\pi/T = 1.4J_0$. In contrast to the $^{39}$K case, in our rubidium studies we work with fixed arrays and instead we vary the interaction-to-tunneling ratio $V/J_0$ by changing the value of $J_0$ (in coordination with $\Delta_0$ and $\omega$). Different initial phases $\varphi_0 = 0, \pi$ were used to realize the opposite sign of potentials, which effectively mimics an interaction sign flip.

All of our measurements are based on selectively de-exciting Rydberg atoms and performing fluorescence imaging of ground state atoms. Site-specific population measurements are attained by applying a series of strong microwave $\pi$ pulses prior to the state-selective de-excitation, after having evolved with the synthetic lattice Hamiltonian for a given duration.

\subsection{Dipole-dipole interactions}

After being excited to Rydberg states, the atoms experience strong dipolar interactions with one another. In our setup, we employ relatively large inter-atom spacing, such that resonant dipolar exchange interactions are dominant. When atoms occupy different states with dipole-allowed matrix elements, they can resonantly exchange their states (excitations) via dipolar exchange~\cite{Browaeys_2016}. Consequently, the interaction strength (or the rate of this exchange) scales with $V_{\text{dd}}/h = -C_3/(2r^3)$, where the value of the $C_3$ coefficient depends on the involved states and $r$ is the interparticle distance. 

In this work, we first focus on the interactions of atom pairs, as illustrated in Fig.~\ref{FIG:fig1} (a). Labeling the atoms in a pair as $1, 2$, the interaction can be formulated as:
\begin{equation}\label{eqn:int}
\begin{split}
    H_{int} = & \sum_{m}\left[ V_{m,\text{intracell}}(a^{\dagger}_{m,1} b^{\dagger}_{m,2}a^{\phantom\dagger}_{m,2} b^{\phantom\dagger}_{m,1} + \rm{h.c.}) + \right. \\& \left. V_{m,\text{intercell}}(b^{\dagger}_{m,1} a^{\dagger}_{m+1,2}b^{\phantom\dagger}_{m,2} a^{\phantom\dagger}_{m+1,1} + \rm{h.c.})\right]
\end{split}
\end{equation}
where $m$ is the index of the unit cell. 

Experimentally, transitions with comparable $C_3$ coefficients were used to construct the lattice for pumping, ensuring that the interactions have an approximate discrete translation symmetry along the lattice. This symmetry enables band structure analysis in the presence of interactions. Indeed, for our large system simulations we assume uniform interactions, $V_{m, \text{intercell}} = V_{m,\text{intracell}} = V$. Due to the sparsity of transitions between states hosting uniform $C_3$ coefficients, combined with our limited microwave bandwidth ($\lesssim 9$~GHz), our experimental implementation is constrained to a lattice of only 5 sites.

The strength of interactions for $^{39}$K atoms was controlled via the inter-atomic spacing, while two distinct sets of transitions were used to alter the sign of interactions. For positive $V$,
$S_{1/2,1/2}$ and $P_{3/2,1/2}$ states (with $L_{J,m_J}$ notation) were used with calculated $C_3$ values $\{-1502.4, -1289.5, -1657.4, -1422.4\}~ \rm{MHz \ \mu m ^3}$; for negative $V$, $S_{1/2,1/2}$ and $P_{3/2,3/2}$ states were used with $C_3$ values $\{1126.8, 967.2, 1243.0, 1066.8\}~ \rm{MHz\ \mu m ^3}$~\cite{sibalic17soft}. To explore the interaction dependence, the separation was tuned from 9.7 to 4.85~$\rm{\mu m}$, resulting in dipolar exchange rates $V/h$ ranging from 0.8 to 6.4~MHz for positive $V$ and 0.6 to 4.8~MHz for negative $V$. We note that as $|V|$ approaches the Rabi rate of the detection pulses, detection fidelity becomes the primary source of discrepancy between ideal theory simulations and experimental data. More discussion on state preparation and measurement (SPAM) errors can be found in Appendix~\ref{appendixB}. 

For $^{87}$Rb, operating with higher $n$ states, the relevant $C_3$ values were $\{-6.30, -5.99, -6.75, -6.41\}~ \rm{GHz\ \mu m ^3}$. A fixed separation of 9.3 $\mu$m was used, giving an average exchange rate of $V/h = 4.0(1)\ \text{MHz}$.
We only use a set of states with positive $V$ for $^{87}$Rb, but the pumping dynamics for negative $V$ are effectively probed by changing the initial pump modulation phase and examining population pumping in the opposite direction (starting from the lattice center).

\section{Two-body pumping}

\begin{figure}[b!]
\includegraphics[width=0.5\textwidth]{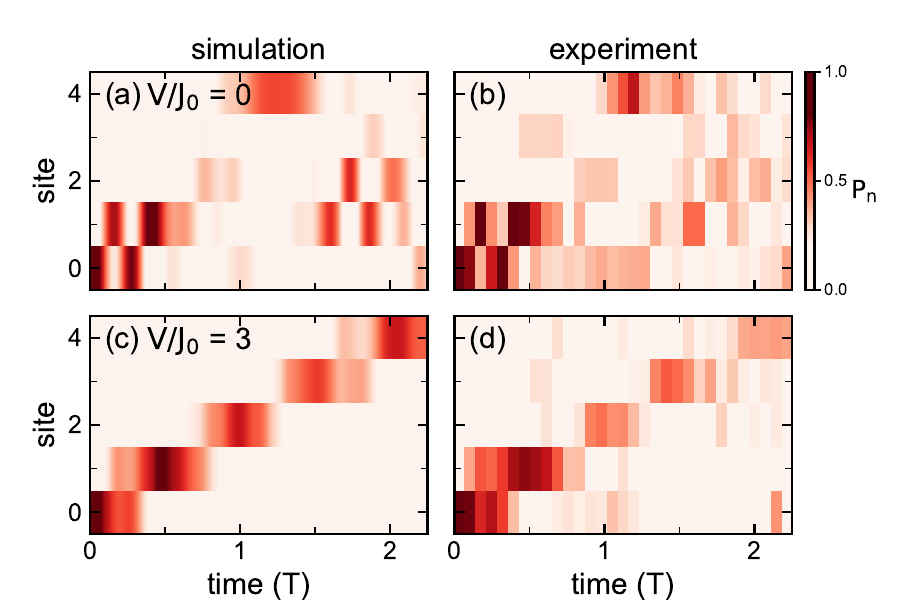}
	\caption{\textbf{Measured pumping dynamics in a 5-site lattice.}
        Pump parameters [$J_0/h = 0.75(1)$~MHz, $\Delta_0/h = 1.5(2)$~MHz, and $\omega = 0.75$~$\rm{MHz}$] are the same as for the 50-site simulations of Fig.~2. Time evolution of the site-wise populations for \textbf{(a,b)} non-interacting and \textbf{(c,d)} a system of interacting atoms. Measurements for $V/J_0 = 0, 3$ are acquired simultaneously and based on post-selection of singly and doubly occupied dimer arrays.
        Numerical simulations use the exact $C_3$ coefficients for each transition and include state preparation and measurement (SPAM) errors. At each time step and site, 100 shots were taken (for arrays of five separated dimers). This corresponds, on average, to $\sim$250 samples for singles and $\sim$150 samples for pairs, giving a statistical uncertainty of $\sim$0.06 for each experimental $P_n$ value.
        }
        
\label{FIG:fig2b}
\end{figure}

We begin by investigating the effect of interactions on the pumping dynamics of atom pairs, using idealized large system simulations (Fig.~\ref{FIG:fig2}) to complement our experimental measurements in a 5-site synthetic lattice (Fig.~\ref{FIG:fig2b}). Both simulations and experiments demonstrate directional transport
accompanied by a reduced spreading of the population distribution envelope, most notably in the presence of intermediate positive-valued interactions. These observations in strongly interacting two-particle systems are highly reminiscent of the behavior of topologically pumped nonlinear photonic fluids~\cite{Jurgensen2021}.

\subsection{Idealized pumping dynamics - theory}

Simulations were conducted on a 50-site lattice, with population initially localized at the central site with the state $|00\rangle$ (i.e., both atoms positioned in the central synthetic lattice site with index $0$). Uniform nearest-neighbor exchange interactions along the synthetic lattice are assumed. Pumping dynamics across the lattice for $V/J_0 = 0, \pm 3$ are shown in Fig.~\pref{FIG:fig2}{a}. Directional transport is observed for all cases, but the interactions lead to qualitative changes in the pumping behavior, including the degree of directional transport and the degree of spreading across the synthetic lattice. These simulations are based upon the experimental pumping trajectory and pumping rate used in the experiments with $^{39}$K.

\begin{figure*}[t]
	\includegraphics[width=0.98\textwidth]{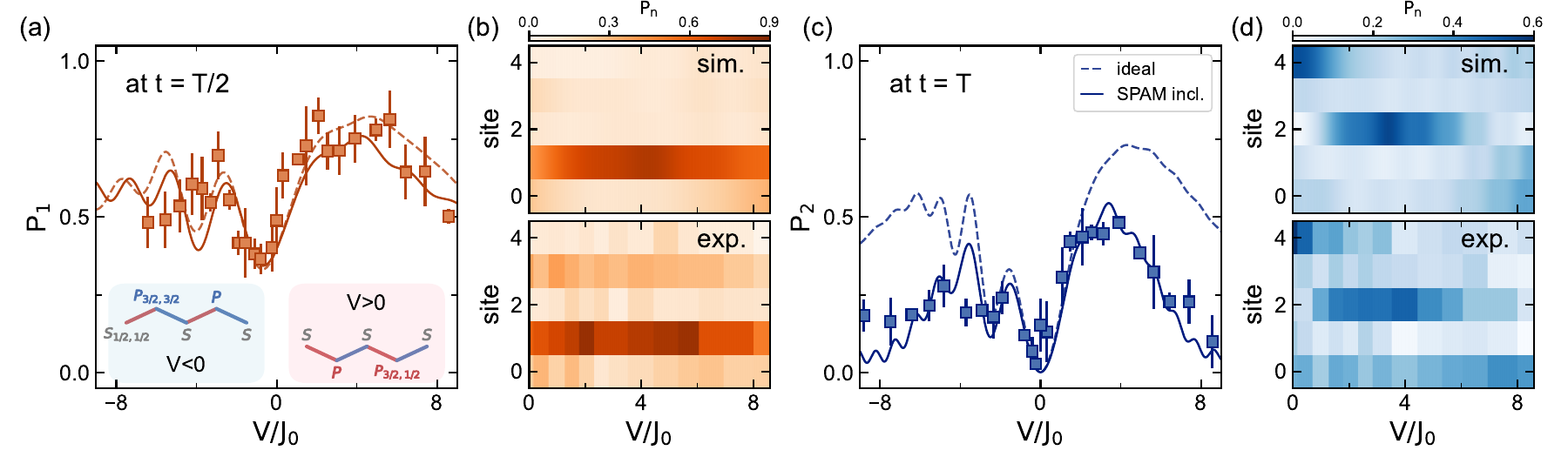}
        \centering
	\caption{\label{FIG:fig3}
        \textbf{Measured interaction dependence of two-body pumping.} 
          Interaction dependence of the probability for atoms to be pumped away from the initial site after one-half \textbf{(a,b)} and one \textbf{(c,d)} pumping period, explored for both positive and negative values of the interaction-to-hopping ratio $V/J_0$ based on utilizing different sets of Rydberg states [inset in \textbf{(a)}]. The site-specific probability for atoms to pump one site away at time $t=T/2$ is presented in \textbf{(a)} and the probability to pump two sites away at time $t=T$ is presented in \textbf{(c)}. Both curves show the asymmetric behavior described in Fig.~\ref{FIG:fig1}. \textbf{(b,d)}~Site-wise population maps at fixed $t=T/2$ and $t = T$ as a function of $V/J_0$, with mean-values of the experimentally measured data (bottom) along with SPAM-included simulation (top). Pump parameters and conditions are the same as in Fig.~\ref{FIG:fig2b}. Error bars are the standard error from multiple independent datasets.
}
\label{FIG:fig3}
\end{figure*}

First, we describe how interactions affect the speed at which the center of mass position, $\lambda$, advances, characterizing the overall pumping efficiency. The solid line in the left panel of Fig.~\pref{FIG:fig2}{b} shows $\lambda$ after four pumping cycles as a function of $V/J_0$, with an enhancement for small positive $V$. Roughly speaking, for small $V/J_0$, this interaction-induced modification of the center-of-mass pumping rate can be understood as the interactions influencing the projection onto the effective bands of the dimensionally extended pump model (e.g., bands as in Fig.~\pref{FIG:fig1}{d}, albeit not well-defined in the presence of interactions). In the ideal case of populating (or filling) a single lattice band with a quantized Chern number, a topological pump should result in atoms being pumped on average by 2 lattice sites (one unit cell) per pumping cycle. Our initial state ($\ket{00}$, with an initial pump phase of $\varphi_0 = \pi$) is not perfectly matched to the Wannier state of a single band, but the presence of weak positive-valued interactions increases this projection and thus enhances the pumping rate. For comparison, the dashed line considers an alternate initialization that matches perfectly to the lower band of the non-interacting pump (it considers $(|0\rangle - |1\rangle)\otimes(|0\rangle - |1\rangle)$ and $\varphi_0 = 3\pi/2$, which is a product state of the single-particle single-band Wannier states at that modulation phase), approaching ideal pumping for $V/J_0 = 0$. Because we assume a fixed pumping rate, the enhancement of the center of mass displacement rate $\dot{\lambda}$ due to enhanced single-band projection for positive interactions is obscured by an eventual breakdown of adiabaticity for very large $|V|$ (detailed in Appendix~\ref{appendixD}). Simply put, the atoms get stuck in place when the pumping rate is too fast for the interaction-slowed dynamics.

This breakdown of adiabaticity for large $|V/J_0|$ hints at a second, somewhat more dominant effect of the dipolar interactions - to inhibit the uncorrelated spreading across the synthetic lattice. In effect, strong exchange interactions can bind the atoms together (by making uncorrelated hopping be non-resonant), forcing them to move only in a coordinated fashion. We can characterize the influence of this dipolar binding by calculating the localization of the atomic distributions along the synthetic dimension through the inverse participation ratio ($\textrm{IPR} = \langle\sum_{i} P^2_i / 2 \rangle$, with $i$ the synthetic site index and taking the average over the two atoms, with $P_i$ the probability to be found at site $i$). We see from the central panel of Fig.~\pref{FIG:fig2}{b} that the IPR grows with increasing $|V|$, more prominently on the side of positive interactions.

Finally, the right panel of Fig.~\pref{FIG:fig2}{b} considers an experimentally convenient heuristic measure, namely the probability to find the atoms $2n$ sites ($n$ unit cells) away from the initial position after $n$ full pumping cycles. This measure, which captures both the propensity of the atoms to pump and their tendency to stay self-localized, shows a marked peak at intermediate positive values of $V/J_0$, along with a smaller peak at negative interaction values.

\subsection{Pair pumping dynamics - experiment}

We now present in Fig.~\ref{FIG:fig2b} the results of the experimental pumping of pairs of atoms. Here, we utilize the aforementioned 5-site lattice and initialize the population at one open boundary.  While the small finite-size nature of this lattice and reflection from the boundaries lead to non-interacting dynamics that are quite distinct from the $V=0$ simulations in an idealized 50-site lattice, such edge effects become negligible when interactions cause the atoms to move in a correlated fashion. Pumping dynamics for $V/J_0 = 0, 3$ are measured out to 3~$\mu$s of evolution in steps of 0.1 $\rm{\mu s}$. Each data point corresponds to 500 experimental dimer samples (100 runs, with 5 independent dimers per run), post-selected based on double occupancy in an initial image. The measured data is presented along with simulations that incorporate separately calibrated errors associated with state preparation and measurement (SPAM), detailed in Appendix~\ref{appendixB}. For $V/J_0 = 0$, the population spreads quite wildly across the lattice, with good qualitative agreement between data and theory. In contrast, for $V/J_0 = 3$, the population tends to stay localized along the synthetic dimension, displaying regular and uniform pumping, in good qualitative agreement with the predictions of the idealized, large-system dynamics.

\begin{figure*}[t!]
	\includegraphics[width= \textwidth]{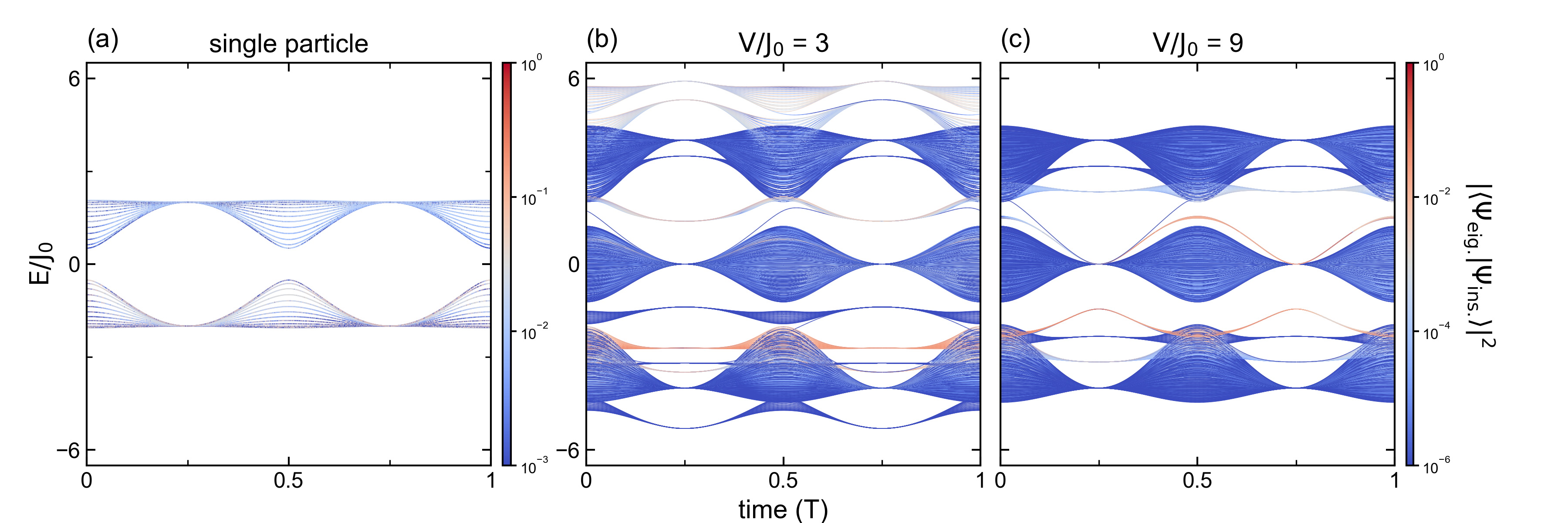}
	\caption{\textbf{Pumping dynamics as projected onto the instantaneous eigenstate energy structure.}
    Simulations are calculated on a 50-site lattice under periodic boundary conditions for 1 period, under initial states $|0\rangle$ [panel (a)] or the pair product state $|00\rangle$ [panels (b,c)]. Colors indicate the overlap between the time-evolved state vector and the instantaneous energy eigenstates of the system Hamiltonian. \textbf{(a)} For the single particle pump, the initial state partially populates both the lower and upper energy bands, with a lower band overlap of $\sum|\langle \Psi_{l}|00\rangle|^2\approx 0.82$. \textbf{(b)} With intermediate interaction $V/J_0 = 3$, the initial two-particle state mainly populates one energy sub-band of the two-atom eigenenergy structure.
    \textbf{(c)}~For an increased interaction strength of $V/J_0 = 9$, the initialized two-particle state again mainly projects onto a single energy sub-band at time $t=0$, but a reduced energy band gap (to symmetry-relevant states) leads population of the time-evolved state to exhibit transfer to another energy sub-band at $t\approx T/4$, signaling a breakdown in adiabaticity. To note, in panel (c) there are additional energy bands centered at energies $\pm 9 J_0$, out of the range of the plot.
    }
\label{FIG:fig4}
\end{figure*}

We now experimentally examine the influence of the interaction strength on the pumping behavior, fixing to specific pumping durations. As with the full pumping dynamics, we find good qualitative agreement in our 5-site lattice with the more idealized behavior of the large-system dynamics. Figure~\ref{FIG:fig3} explores the interaction dependence of the pumping behavior at two specific time points, $t = T/2$ and $T$. As we display in the lower inset of Fig.~\pref{FIG:fig3}{a}, alternate sets of Rydberg levels are used to encode the synthetic lattice sites when exploring the cases of negative and positive $V$.

In the panels Figs.~\pref{FIG:fig3}{a,c}, we present the measured probability $P_{2n}$ for atoms to be pumped over by $2n$ sites after $n$ pumping cycles, plotting the measured population $P_1$ after a half pump cycle in panel (a) and the population $P_2$ after a full pump cycle in panel (c). In rough qualitative agreement with the idealized behavior of Fig.~\pref{FIG:fig2}{b}, we observe that both positive and negative interactions lead to enhanced $P_{2n}$ values, with a more pronounced enhancement for positive $V/J_0$ and some suppression of enhancement at large $|V| \gg J_0$ due to a breakdown of adiabaticity. We note one conspicuous feature of the data in panel (c) - the measured suppression of $P_2$ for large values of $|V|$ is much larger than what we naively expect based on ``ideal'' simulations (dashed lines). As our measurement of the state $\ket{2}\equiv \ket{43S_{1/2,1/2}}$ necessitates a pair of shuttling microwave $\pi$ pulses to the de-excited $\ket{42S_{1/2,1/2}}$ state prior to readout, very large interactions inhibit this state transfer process due to limited Rabi rates of the shuttling microwave pulses.  In Figs.~\pref{FIG:fig3}{a,c}, we additionally show solid ``SPAM-included'' theory lines that account for state preparation and measurement error, specifically by modeling the actual influence of interactions during the state readout shuttling process. For large $|V|$, these ``SPAM-included'' simulation curves find much better agreement with the experimental data.

Finally, for positive interactions, we demonstrate the more comprehensive agreement between our experimental measurements and theory (including the influence of interactions during readout) considering the interaction dependence of all site populations after a half pumping cycle [Fig.~\pref{FIG:fig3}{b}] and a full pumping cycle [Fig.~\pref{FIG:fig3}{d}]. To summarize, our experimental data on the interaction-dependence of pair-pumping behavior in a five-site synthetic lattice reflects the main qualitative features as predicted from an idealized large-system behavior: a peak in the enhancement of the pumping probability for intermediate $|V/J_0|$ values and an asymmetric dependence with respect to the sign of the interactions.

\subsection{Pairwise pumping mechanism}

The Thouless pumping behavior of single atoms is well-defined by the topology of the energy bands, characterized by the Chern number, under dimensional extension~\cite{Thouless1983,citroThoulessPumpingTopology2023}. However, the presence of interactions complicates the determination and the very definition of the Chern number. In this context, we present the analysis of instantaneous many-body spectra under the experimental pumping trajectories to elucidate the mechanisms underlying our observation of interaction-enhanced pumping and interaction-inhibited spreading of population across the Rydberg synthetic lattice. Additionally, for the simple case of two particles in the large interaction limit, we can derive an effective description based on second-order perturbation theory in which particles are restricted to undergo modified, but still-topological, pumping in a restricted basis of bound pair states.

We first present results on the evolution of the adiabatic (or instantaneous) eigenenergy spectra over the pumping cycle (based on 50-site lattices with periodic boundary conditions) for (a) single particles, (b) pairs of particles under intermediate interactions $V/J_0 = 3$, and (c) pairs under strong interactions $V/J_0 = 9$. For the single-particle scenario in Fig.~\pref{FIG:fig4}{a}, one observes the simple two-band spectrum of the dimensionally-extended Rice-Mele model as presented earlier in Fig.~\pref{FIG:fig1}{d}. The plots feature a color scale that evolves with time $t$, representing the overlap of the time-evolved state with the pump's instantaneous eigenstates (assuming the same initial state and pump trajectory as in experiment and Fig.~\ref{FIG:fig2}). In the single-particle case, this projection is primarily onto the lower band and shows negligible transfer between the bands. As discussed before, the initial projection onto the lower band is incomplete due to imperfect overlap of the $\ket{0}$ state with the lower-band Wannier state.

\begin{figure}[t!]
	\includegraphics[width=0.5 \textwidth]{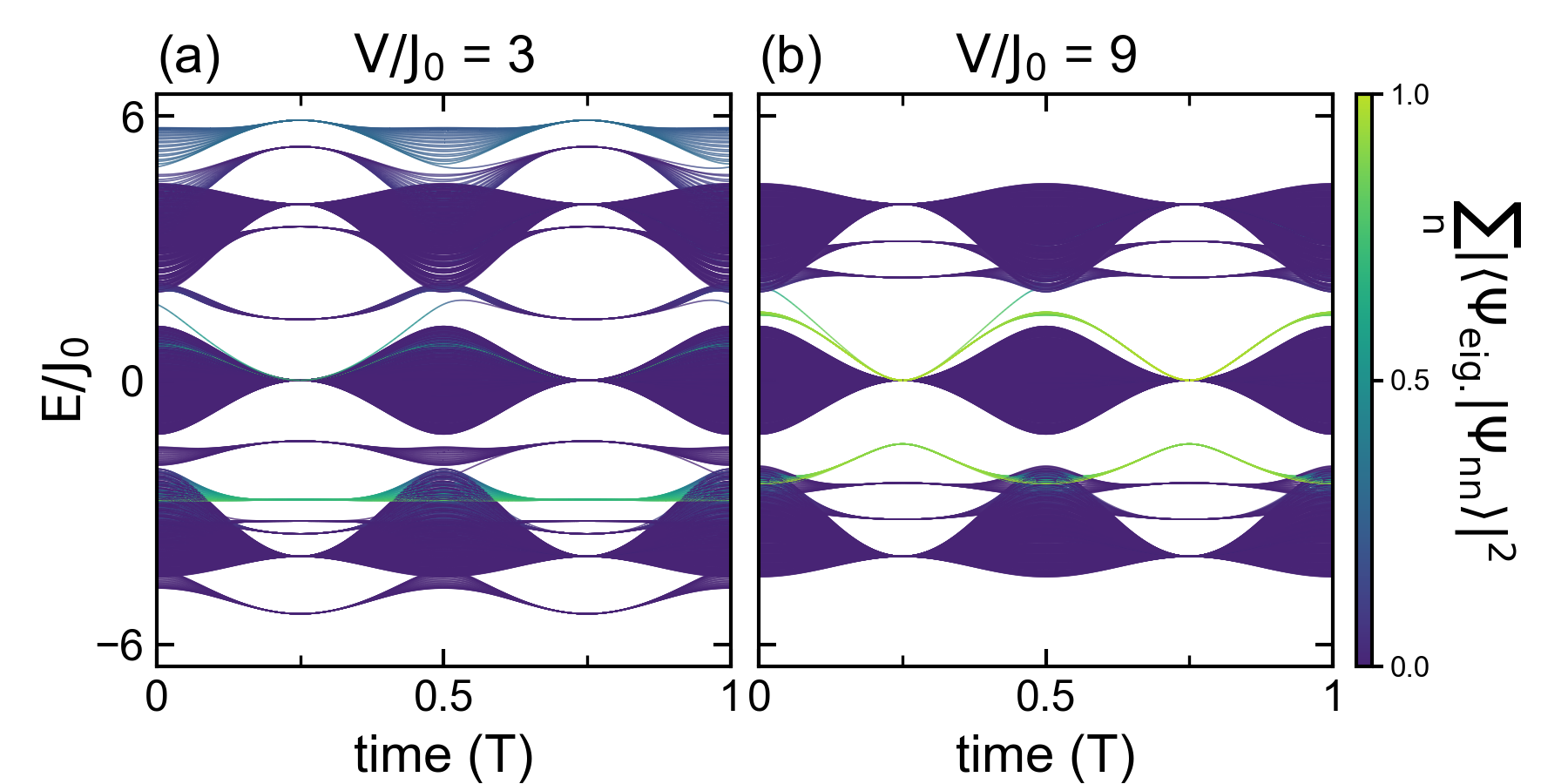}
	\caption{\textbf{Pair-state analysis of two-particle energy bands.}
    The two-particle energy spectra are the same as in Fig.~\pref{FIG:fig4}{b,c}, but here the color indicates the summed overlap of the instantaneous eigenstates with all basis states of the form $|nn\rangle$ (i.e., two atoms residing in the same internal state).
    Note that, for $V/J_0 = 9$, there are additional energy eigenvalues lying outside of the range of the plot.
    }
\label{FIG:fig5}
\end{figure}

Figures~\pref{FIG:fig4}{b,c} display the evolution of the analogous two-particle spectra for the cases of intermediate ($V/J_0 = 3$) and large ($V/J_0 = 9$) interactions. One key feature of these spectra is the appearance of (bands of) two-particle eigenmodes centered around energies of $\pm V$ due to the exchange interactions. The bands of such states lie outside of the range of energies plotted in Fig.~\pref{FIG:fig4}{c}, appearing at energies $\sim \pm V = \pm 9 J_0$.
In the large interaction limit $V \gg J_0$, the interaction-free initial state $\ket{00}$ will primarily project onto the central ``non-interacting'' bands of two-particle eigenstates. As seen in Fig.~\pref{FIG:fig4}{b}, for $V/J_0 = 3$, the initial projection is primarily onto a specific central band of states and the time-evolved state remains within this singular band. For large interactions, Fig.~\pref{FIG:fig4}{c} for $V/J_0 = 9$, population remains within such a non-interacting set of states but undergoes transitions between two sub-bands due to a breakdown in adiabaticity for the considered pump rate.

To gain further insight into the structure of the two-particle energy bands, we explore in Fig.~\ref{FIG:fig5} the projection of the instantaneous eigenstates onto a sub-basis of states in which the two atoms reside on the same synthetic lattice site, $\ket{nn}$, where $n$ is the site index. The initial state $\ket{00}$ will have full overlap with such states, and we furthermore expect that the time-evolved state [as studied in Figs.~\pref{FIG:fig4}{b,c}] will retain overlap when $V \gg J_0$ due to the inhibition of uncorrelated hopping.
In comparing Fig.~\ref{FIG:fig5} and Fig.~\ref{FIG:fig4}, we observe that the bands with which the time-evolved state has sizable projection are also the bands with pair-wise occupancy of the atoms on the same internal state (i.e., ``pair states'' of the form $\ket{nn}$). 
In other words, the particle is pumped along a truncated state basis consisting of all $\ket{nn}$ states.
In the case of $V = 9 J_0$, where we observed that population transferred between two bands in Fig.~\pref{FIG:fig4}{c}, we see from Fig.~\pref{FIG:fig5}{b} that both of these bands have full projection onto the pair state basis of $\ket{nn}$ states. That is, the particles remain within the truncated basis of pair states, but exhibit a breakdown of topological pumping.

\begin{figure*}[t!]
	\includegraphics[width= \textwidth]{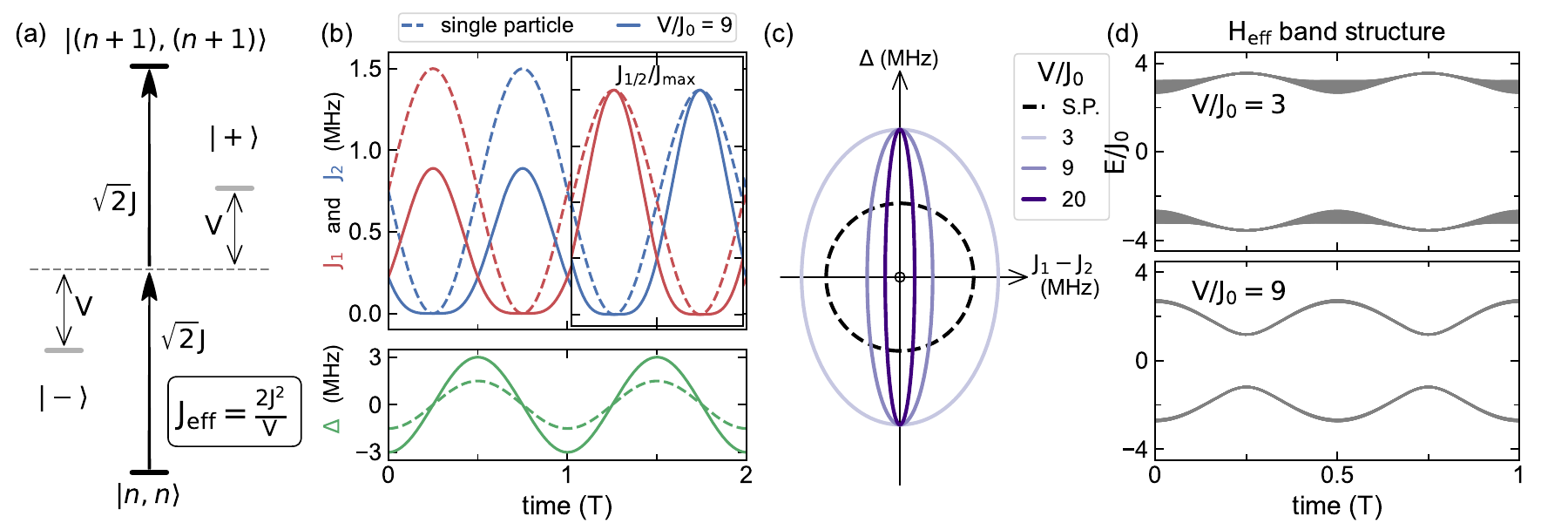}
	\caption{\textbf{Effective model for pair-state topological pumping.} \textbf{(a)} A depiction of the effective second-order process for a pair of atoms to transition between adjacent pair states $\ket{n,n}$ and $\ket{(n+1),(n+1)}$ (assuming a single-particle hopping $J$ and zero site-energy imbalance, for simplicity). The microwave-induced hopping rate from pair product states to entangled triplet states of the form $\ket{+} = (\ket{n,(n+1)}+\ket{(n+1),n})/\sqrt{2}$ is enhanced from the single-particle case by a factor of $\sqrt{2}$. Assuming a global, uniform microwave drive, there is no matrix element connecting pair product states to the anti-symmetric singlet state $\ket{+} = (\ket{n,(n+1)}-\ket{(n+1),n})/\sqrt{2}$. Due to dipolar interactions, both the triplet and singlet states are shifted away from the non-interacting resonance condition by a binding energy $V$. In the limit of large interactions, $V \gg J$, atoms will undergo an effective pair-state hopping at a rate $J_\textrm{eff} = 2 J^2/V$.
    \textbf{(b)} A comparison of the time-dependent parameter modulation of the single-particle Rice-Mele model (dashed lines) and the effective parameters of the pair-state Rice-Mele model. The inter-site bias modulation $\Delta$ is simply doubled for pair product states, and the effective hopping terms are modified as suggested in panel (a). In addition to the change in their maximum values, we can see from the normalized parameter modulations appearing in the inset box in the upper right that the shape of the path in the $\{\Delta,J_1-J_2 \}$ parameter space will be modified as well.
    \textbf{(c)}~Pumping trajectories in the Rice-Mele model parameter space $\{\Delta,J_1-J_2 \}$ for various $V/J_0$ values. The black dashed line indicates the single-particle case. For the pair-state model, pumping trajectories remain topological (encircling the origin), but become increasingly compressed along the $J_1-J_2$ axis for increasing $V$ due to the scaling of the pair state hopping rate [as $\sim 2J(J/V)$ for fixed $J$].
    \textbf{(d)}~Instantaneous bands for the effective pair-state Rice-Mele model, shown for $V/J_0 = 3$ and 9. The bandwidth is compressed due to the scaling of $J_\textrm{eff}$, and likewise the overall compression of the pumping trajectory along the $J_1-J_2$ parameter axis leads to a decrease of the minimal gap (at times $T/4$ and $3T/4$) of the instantaneous spectrum.} 
 
\label{FIG:fig6}
\end{figure*}

The above observations based on the two-particle energy spectra further suggest an effective pair-state model describing the pumping of atoms initialized in states such as $\ket{00}$ in the large interaction limit (where interactions prohibit particles from leaving this pair subspace). Figure~\ref{FIG:fig6} presents the basic motivation behind this effective model for pairwise pumping. The states of this model are the pair states $\ket{nn}$, which have diagonal energy terms $\Delta_\textrm{eff}$ that are simply twice that of the single particle model of Eq.~\ref{eqn:sRM}. The off-diagonal terms representing hopping between such pair states scale as $2 J_0^2/V$, and thus give rise to a much narrower (in energy) band of states in the $V \gg J_0$ limit. Importantly, the effective off-diagonal couplings between pair states retain the staggered $J_1$-$J_2$ form of the parent single-particle model (Eq.~\ref{eqn:sRM}).

While the resulting pair state Rice-Mele model remains topological, it is found to differ from the initial single-particle model in one important way. Whereas the diagonal energy imbalance $\Delta_\textrm{eff}$ between adjacent pair states still varies sinusoidally in time, the corresponding difference between adjacent off-diagonal coupling terms, $\delta J = J_\textrm{1} - J_\textrm{2}$, changes more abruptly than in the single-particle model, as shown in Fig.~\pref{FIG:fig6}{b}. These modified trajectories of the effective model parameters, as depicted in Fig.~\pref{FIG:fig6}{c}, cause the effective Rice-Mele model to become more akin to the ``control freak''~\cite{Asbóth2016} protocol as the interactions are increased. In the extreme ``control freak'' limit of a topological pump, population pumps along a lattice without any spreading due to a negligible bandwidth. Finally, as seen in Fig.~\pref{FIG:fig6}{d}, the bandgap of the effective pair-state model decreases with increasing $V$, leading to an eventual breakdown of adiabaticity. This breakdown leads to the transfer of population between energy sub-bands as observed in Fig.~\pref{FIG:fig4}{c} and Fig.~\pref{FIG:fig5}{b}. Furthermore, it underlies our experimental observation of a decreased downstream pumping efficiency for large $|V| \gtrsim 6$ in Fig.~\pref{FIG:fig3}{c} (this intrinsic effect of the model system is separate from the additional influence of interactions on our state readout process). In Appendix~\ref{appendixE} we consider how, for two particles in the limit of large interactions (for $V/J_0 = 9$), adiabaticity can be restored by operating at lower pump rates.

An alternative picture for the emergence of the effective pair state Rice-Mele model can be seen by considering the Fock-space of two interacting atoms. In this representation, the Hamiltonian for two interacting particles in one dimension can be considered as a non-interacting problem in two effective dimensions. Without interactions, one finds topological pumping to appear separably along both axes of this two-dimensional system, relating to independent pumping of the particles. Dipolar exchange interactions will lead to off-diagonal matrix elements adjacent to the diagonal ``sites'' in this two-dimensional space. When the dipolar interactions are very large, $V \gg J_0$, the dynamics of pair states become confined along the diagonal in Fock space, reminiscent of the picture for emergent correlated pair-wise hopping in Hubbard systems~\cite{Preiss-QWalk,joyce}.

\section{Few- and many-atom pumping}

The experiments and numerical results of the preceding Section suggest a complete and simple picture for the topological pumping of dipolar-bound atoms in the two-body limit, understandable through a microscopic description in terms of emergent topological pumping in a pair-state basis. We now move on to consider how dipolar exchange interactions modify the internal state pumping dynamics in few- and many-atom arrays.

For just two atoms, we saw that dipolar interactions led to the topological pumping of bound pairs, with some qualitative similarity to collective pumping dynamics observed in classical nonlinear photonics~\cite{Jurgensen2021}. Extending to arrays with higher atom numbers, both for all-to-all connectivities in space (for three atoms in triangular arrays) and for graphs with non-uniform but long-ranged spatial connectivities, we seek to further explore how interactions modify the topological pumping dynamics of Rydberg atoms along a synthetic dimension.

\begin{figure}[t!]
	\includegraphics[width=0.5\textwidth]{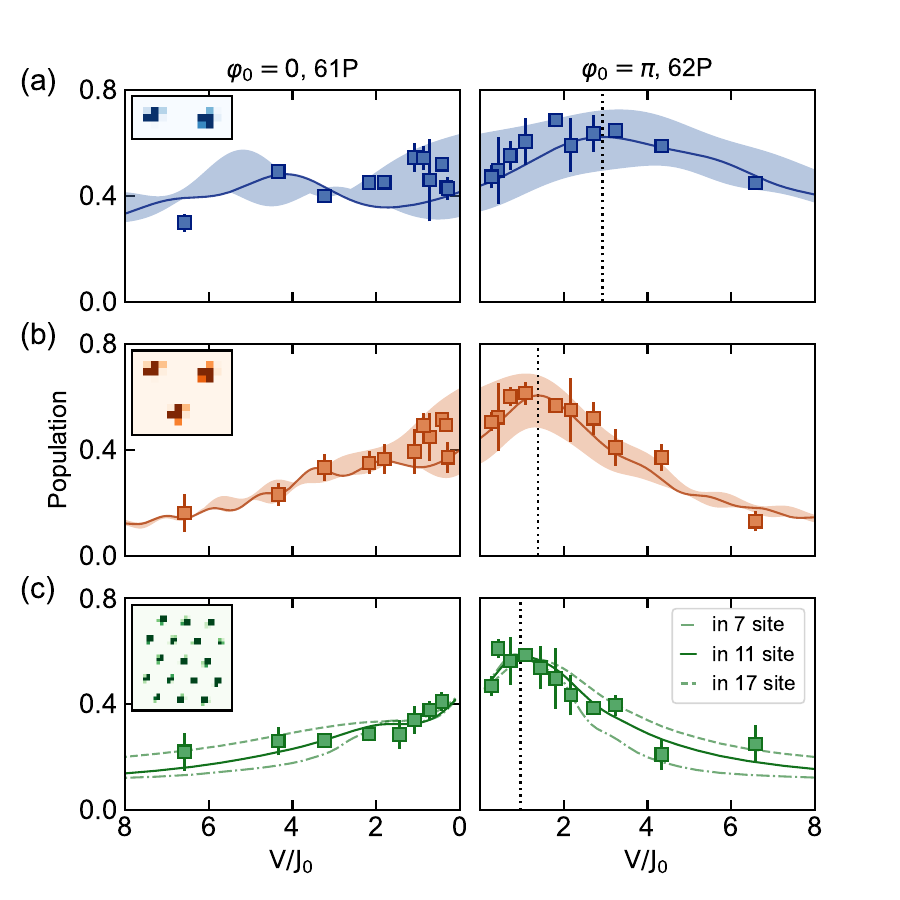}
        \centering
	\caption{\label{FIG:fig7 ?}
\textbf{Measured interaction dependence of the pumping behavior with different real-space geometries.} Three different real-space patterns are used to explore the few and many-atom pumping, \textbf{(a)}~dimers, \textbf{(b)}~equilateral triangles, and \textbf{(c)}~17-site triangular lattices (all with nearest neighbor spacings between atoms of 9.3~$\mu$m). For dimer and triangle patterns, we consider fully loaded samples by post-selecting on the first image, while for the lattice, we post-select shots with more than 75\% loading. For all the patterns, the nearest neighbor distance is the same, fixing the nearest-neighbor exchange interaction to a value of $V/h = 4.0(1)$~MHz. To measure the dependence on the interaction-to-tunneling ratio, $V/J_0$, we vary the tunneling strength $J_0$. All the data here are taken under the same pumping conditions, where $\Delta_0 = 2.8 J_0$ and $\omega = 2\pi/T = 1.4 J_0$. The modulation time is rescaled with respect to $J_0$ to ensure a fixed $t = 0.64T$.
In \textbf{(a,b)}, solid lines correspond to the simulation with SPAM errors, with the shaded bands indicating the uncertainties from experimental parameters. Due to an inability to directly simulate the dynamics of $\sim$13 interacting five-level atoms, the theory lines in \textbf{(c)} are for seven atoms filling arrays of varying size (uniformly filling seven sites, randomly filling eleven sites, or randomly filling a seventeen sites, as described in the main text), meant to capture the effect of varying particle density. For the partial filling cases, theory curves are averaged over 100 random loading patterns.
Error bars are the standard error from multiple independent datasets.
}
\label{FIG:fig7}
\end{figure}

\subsection{Experimental pumping and interaction dependence for different real-space geometries}
As in the preceding two-atom studies, we implement a Rice-Mele topological pump in the internal state space of Rydberg atoms. To enable the extension to many-atom arrays, we utilize a separate apparatus for $^{87}$Rb that allows for atom trapping in two dimensions. Incidentally, this system also allows for excitation to higher-lying Rydberg levels (due to electric field cancellation). So, as described before, the five-state ``lattice'' utilized for these studies involves $S$ and $P$ levels with $n \in \{ 61,62,63\}$.

Figure~\ref{FIG:fig7} shows the dependence of the probability for Rydberg electrons to be pumped along the synthetic dimension as a function of the dipolar interaction-to-microwave-driven hopping ratio $V/J_0$, for several different geometries of the atoms in space (similar to Fig.~\ref{FIG:fig3}). We first describe the few-atom case, where we reproduce in (a) the dynamics of just two interacting atoms and also extend in (b) to an arrangement of three atoms on a triangular grid with uniform pairwise interactions. We first describe the few-atom case, where Fig.~\pref{FIG:fig7}{a} reproduces the dynamics of two interacting atoms and Fig.~\pref{FIG:fig7}{b} extends to an arrangement of three atoms in a triangle with uniform pairwise interactions.
We begin with all atoms in the Rydberg level $62S_{1/2}$ ($m_J = 1/2$) and look at the population arriving one synthetic site away after half a pumping period. Here, we operate with a fixed arrangement of states and thus a fixed sign of the dipolar exchange interaction, $V > 0$. Positive values of $V/J_0$ are probed in the same way as in Fig.~\ref{FIG:fig3}, by starting with a pump phase of $\varphi_0 = \pi$ and measuring the final population in a state further up the internal state lattice, namely $62P_{3/2}$. This positive interaction data appears in the right panel of Fig.~\ref{FIG:fig7}. We can effectively explore negative $V/J$ values by instead starting with a pump phase of $\varphi_0 = 0$ and measuring the population that appears displaced in the opposite direction along the synthetic lattice, namely $61P_{3/2}$. This negative interaction data appears in the right panel of Fig.~\ref{FIG:fig7}. Because we explore pumping in both directions along our few-site synthetic lattice, here we begin with population at the central synthetic lattice site (whereas in the two-body studies on $^{39}$K we began at the end of our small Rice-Mele lattices).

Qualitatively, we observe the same overall trend as in Fig.~\ref{FIG:fig3} - a peak in the ``downstream pumping efficiency'' appearing at moderate positive values of $V/J$. 
For the three-atom case, we find the same qualitative trend as for two atoms, although the peak in the pumping efficiency appears at a smaller value of $V/J$, roughly half that of the two-atom case.
Both of these sets of data are in good agreement with the full dynamics simulations (solid lines, with shaded regions incorporating parameter uncertainty and SPAM). We note that, because the data of Fig.~\ref{FIG:fig7} is based on a slightly different pumping trajectory (and rate) in parameter space as compared to that of Fig.~\ref{FIG:fig3}, the exact trend of the two-atom dynamics is slightly different. Most notably, the smaller peak appearing at negative $V/J$ values in Fig.~\ref{FIG:fig3} is nearly absent in Fig.~\ref{FIG:fig7}, consistent with the simulations under this slightly modified pumping.

In Fig.~\pref{FIG:fig7}{c}, we explore how the interaction-dependent pumping behavior changes when we move from all-to-all interactions to an extended two-dimensional graph. We consider atoms probabilistically loaded into a high-connectivity triangular array as depicted in the inset. In practice, we load atoms into a rectangular 5~$\times$~7 array with asymmetric spacings along the vertical and horizontal axes (with a ratio of $\sqrt{3}$). An additional set of ``removal'' tweezers induces loss at half of the sites, leading to an effective probabilistic loading of a triangular array. The three-atom pattern in Fig.~\pref{FIG:fig7}{b} is similarly distilled from a 2~$\times$~3 array. The typical atom loading for the data in Fig.~\pref{FIG:fig7}{c} is 80\%, based on enhanced ($> 50\%$) loading and post-selection on shots with 12 or more atoms.

In looking at the interaction dependence of the pumping probability in Fig.~\pref{FIG:fig7}{c}, we observe a continuation of the trend as seen when extending from two to three atoms. The peak of the probability to be found one site displaced after half a pump period moves in towards still smaller values of $V/J_0 \approx 1$. Additionally, the drop-off in pumped population for larger $V/J$ is steeper than in the two- and three-atom cases, suggesting a decreased mobility in the presence of large interactions.

An exact simulation of the array dynamics, involving approximately 13 five-level atoms, is largely intractable. Instead, we compare the triangular array data to several sets of simulations based on a fixed number of seven five-level atoms filling arrays at varying densities. Specifically, we consider a unit-filling seven-site array (six atoms surrounding a central site, dashed-dotted line), seven atoms randomly filling an array of 11 sites (central three rows, solid line), and seven atoms randomly filling the experimental 17-site array (dashed line). Although none of these simulations should be seen as an exact proxy for the true experimental arrangement, the experimental data seem to be in best agreement with the intermediate case of 7 atoms filling 11 sites (averaging over 100 different random configurations). The atom density of this simulation is less than that (postselected upon) for the data, however the Rydberg atom density in our arrays is slightly lower than the initial ground state atom density due to imperfect STIRAP excitation probability.

\subsection{Dependence on geometry and connection to mean-field soliton pumping}

Overall, the few- and many-atom pumping data in Fig.~\ref{FIG:fig7 ?} agreed well with the full quantum dynamics simulations based on the exact (for few atoms) and approximated (for arrays) atom geometries. Here, we provide some simple physical arguments to better understand the main qualitative trends observed in Fig.~\ref{FIG:fig7 ?}, also providing arguments for why strongly interacting Rydberg electrons pumped along a synthetic dimension would bear any resemblance to nonlinear photonic pumps~\cite{Jurgensen2021}.

A key feature of the data and simulations in Fig.~\ref{FIG:fig7 ?} is the shift of the ``peak'' $V/J_0$ ratio (at which the downstream pumping is maximized) for different geometries. Qualitatively, this trend is consistent with the optimal pumping occurring at a roughly fixed ratio of the mean per-particle interaction (and not simply $V/J_0$). In going from dimer to triangles the interaction per particle doubles, and the optimal $V/J_0$ value is lowered by roughly a factor of two. 
The peak $V/J_0$ is still lower for the array data. While the per-particle interaction in this scenario is less well-defined than for the all-to-all interacting dimers and triangles, the per-atom connectivity is still clearly enhanced in the arrays.

We note that a mean-field description of the behavior of our arrays is not entirely unrealistic, as dipolar-interacting spins in two dimensions should have some mean-field-described properties due to their long-ranged interactions~\cite{Tromb-RMP}. Moreover, spatial frustration should not play a role in these studies, which utilize only global microwave driving from initial product states of the Rydberg electrons. Appendix~\ref{appendixF} explores this (over)simplified mean-field picture through numerical simulations, showing an approximate collapse (when scaled by the per-atom connectivity) of the pumping trends (vs. $V/J_0$) all-to-all-connected quantum pumps.

The apparent similarities between the pumping of Rydberg electrons and optical solitons are at first glance somewhat surprising based on the specific forms of interaction in these two systems. Rydberg electrons experience flip-flop exchange interactions along their synthetic lattice, whereas photonic ``fluids'' experience Kerr nonlinearities that are site-local along their corresponding waveguide array. Here, we give a few arguments for how the similarity of the dynamics observed in these disparate systems can be understood.

We can first understand that these two systems will behave the same in the extreme limits of weak ($|V/J_0| \rightarrow 0$) and strong ($|V/J_0| \rightarrow \infty$) interactions.
In the former limit, the details of the interactions are entirely inconsequential, whereas in the latter limit the interactions will effectively arrest all dynamics in both systems. Second, near the onset of some form of interaction-induced immobilization, the dynamics of collective pumping along the internal dimension can be understood as a successive set of transitions between polarized states on adjacent sites of the lattice. Confining to just two sites along the topological pump, one can observe that the Ising-like (or Potts-like when considering many sites) Kerr nonlinearity of the photonics problem maps onto an infinite range transverse field Ising model. While the dipolar exchange interaction is fundamentally different, when viewed in a mean-field picture (tracing over the real-space positions of the atoms such that $S^i_{x}S^j_{x} \rightarrow S_x^2$) and restricting to two synthetic sites, it will effectively give rise to the same dynamics (as $S_x^2 + S_y^2 = S^2 - S_z^2$). Thus, if one views slow topological soliton pumping as performing a series of adiabatic sweeps in successive two-state infinite-range transverse field Ising models, an analogous process of successive population-inverting transitions appearing in multi-level magnets~\cite{Sundar2018,SundarPRA,Membranes,Sohail-finiteT} is less surprising.

\section{Conclusion}

In this paper, we have investigated the dynamics of Rydberg electrons subject to topological pumping along a synthetic dimension of internal states (cf. Ref.~\cite{Trautmann2024}). In particular, we have examined how dipolar interactions between atoms greatly modify the pumping dynamics, leading to the emergence of multi-particle states that are self-bound (along the internal dimension) that continue to undergo topological pumping. For two atoms, we concretely relate the dynamics to the emergence of an effective pair-state Rice-Mele model that inhibits spreading but maintains topological pumping. We then extend our measurements to systems of three- and more atoms in two-dimensional arrays, observing a consistent picture of interaction-enhanced probability for atoms to appear at a given site along the pumping path, a metric that accounts for both an ability to pump and a propensity to remain self-bound by interactions. As the particle number, or perhaps more importantly the average connectivity of the atoms, grows, we find that this interaction-enhanced pumping of population along the lattice is stabilized at smaller and smaller interaction strengths.

Our observations of topologically pumped self-bound few- and many-body states bear a striking similarity to pioneering results from classical nonlinear photonics~\cite{Jurgensen2021}, and we discussed physical arguments supporting the connection between these systems. In the future, related connections may allow us to translate phenomena from the regime of nonlinear systems to many-body spin arrays, e.g., for applications in quantum state engineering and quantum sensing. Importantly, such connections may afford some intuition for the dynamical behavior of complex many-body spin systems that are otherwise intractable to numerical methods.

Our experiments further demonstrate the utility of performing synthetic dimensions~\cite{Hazz-SynthDim-Rev,Fabre_2024,Arguello-Luengo2024,Yuan-2025} experiments in strongly interacting spin arrays~\cite{Sundar2018}, which allow for a fine control over internal state synthetic lattices and enable well-resolved measurements of both real- and internal-space dynamics. Utilizing only global control, we have shown how this system can be applied to problems in few- and many-body topology, complementing work in real-space optical superlattices~\cite{Walter2023,Viebahn2024}. This approach can be extended to explore the intriguing phenomenon of interaction-enabled topological pumping~\cite{Viebahn2024} in the few-body limit, and can be combined with internal-space gauge fields~\cite{Chen2024,Chen2025} to study few-body quantum Hall systems~\cite{Tai2017,Leonard2023,Jochim-FQH}.
Alternatively, one can begin with the rich physics associated with many-atom arrays of dipolar magnets~\cite{Sundar2018,SundarPRA,Membranes,Sohail-finiteT,Wang2025} and ask what new phenomena emerge from the addition of complex internal state ``synthetic lattices.''

\section*{Acknowledgments}
We thank Mikael Rechtsman, Marius J{\"u}rgensen, and Rhine Samajdar for fruitful discussions, and we thank Tabor Electronics for the use of an arbitrary waveform generator demo unit. This material is based upon work supported by the AFOSR MURI program under agreement number FA9550-22-1-0339 and by the National Science Foundation under grants No.~1945031 and No.~2438226.
K.R.A.H.'s work is supported in part by National Science Foundation (PHY-1848304) and W. M. Keck Foundation (Grant No. 995764).

\appendix

\section{Time-dependent control of microwave (MW) tones}\label{appendixA}
The internal-space, synthetic lattice Hamiltonian is derived from the interaction of the atoms with a time-dependent, multi-tone microwave field. Here, we derive the time-dependent control and its relation to the synthetic lattice Hamiltonian, based on the instantaneous spectrum applied to the atoms.

To implement the time-dependent Rice-Mele Hamiltonian, we apply both amplitude and phase modulations to a series of microwave (MW), as described in earlier works~\cite{Chen2024,Chen2025}. The microwave tones address electric dipole-allowed transitions that couple adjacent pairs of synthetic sites (Rydberg states). We consider a dimerized chain of such sites (with unit cell index $j$), which in the rotating wave approximation is governed by the Hamiltonian $H=H_0 + H_1$ with
\begin{equation}
    H_0 = \sum_j \left(\hbar\omega_{j}^{a}a_j^\dagger a_j^{\phantom\dagger} + \hbar\omega_j^b b_j^\dagger b_j^{\phantom\dagger}\right)
\end{equation}
and
\begin{equation}
\begin{split}
    H_1 = & \sum_j\left[ \frac{\hbar\Omega_j^{ab}(t)}{2}e^{i[\omega_j^{ab}t + \phi_j^{ab}(t)]} a_j^\dagger b_j^{\phantom\dagger} \right. \\  &\left. \ \ \ \ \ \ + \frac{\hbar\Omega_j^{ba}(t)}{2}e^{i[\omega_j^{ba}t + \phi_j^{ba}(t)]} b_j^\dagger a_{j+1}^{\phantom\dagger} \right] + {\rm h.c.} \ .
\end{split}
\end{equation}
The creation (annihilation) operator for even sites $\ket{a, j}$ is $a_j^\dagger~(a_j^{\phantom\dagger})$, while $b_j^\dagger~(b_j^{\phantom\dagger})$ is the corresponding operator for odd sites $\ket{b, j}$. Here $\Omega_j^{ab}(t)= d_j^{ab}\cdot E_j^{ab}(t)$ is the Rabi frequency for the intracell coupling $\ket{a,j}\leftrightarrow\ket{b,j}$ (resonant transition frequency $\omega_j^{ab}=\omega_j^b - \omega_j^a$, dipole moment $d_j^{ab}$) driven by the oscillating MW electric field $\mathcal{E}_j^{ab}=E_j^{ab}(t)\cos{[\omega_j^{ab}t + \phi_j^{ab}(t)]}$ with the (slowly) time-dependent amplitude $E_j^{ab}(t)$ and phase $\phi_j^{ab}(t)$, while the MW tone $\mathcal{E}_j^{ba}=E_j^{ba}(t)\cos{[\omega_j^{ba}t + \phi_j^{ba}(t)]}$ accounts for the intercell coupling $\ket{b,j}\leftrightarrow\ket{a, j+1}$ (resonant frequency $\omega_j^{ba}=\omega_{j+1}^a - \omega_j^b$, dipole moment $d_j^{ba}$) with the Rabi frequency $\Omega_j^{ab}(t)= d_j^{ba}\cdot E_j^{ba}(t)$. In the following, we show how to engineer the exact form of the Rice-Mele model via the flexible programmability over the time-dependent amplitudes and phases for each transition, as realized in our experiment with an arbitrary waveform generator.

We first apply a unitary transformation and arrive at the Hamiltonian under the interaction picture,
\begin{equation}
\begin{split}
    H_I = & e^{iH_0t/\hbar} H_1 e^{-iH_0t/\hbar} \\
     = & \sum_j\left[\frac{\hbar\Omega_j^{ab}(t)}{2}e^{i\phi_j^{ab}(t)} a_j^\dagger b_j^{\phantom\dagger} + \right. \\ & \left. \frac{\hbar\Omega_j^{ba}(t)}{2}e^{i\phi_j^{ba}(t)} b_j^\dagger a_{j+1}^{\phantom\dagger}\right] + {\rm h.c.}~~.
\end{split}
\end{equation}
Then, by applying the gauge transformation $a_j = e^{-i\phi_j^a(t)}\tilde{a_i}$, $b_j = e^{-i\phi_j^b(t)}\tilde{b_i}$ and letting $\phi_j^{ab}(t) = \phi_j^b(t)-\phi_j^a(t)$, $\phi_j^{ba}(t) = \phi_{j+1}^a(t)-\phi_j^b(t)$, we obtain the Heisenberg equation of motion for operator $\tilde{a}_j$ as
\begin{eqnarray}
 i\hbar\frac{d \tilde{a}_j}{dt} & = & -\hbar\frac{d\phi_j^a(t)}{dt}\tilde{a}_j + e^{i\phi_j^a(t)}\left[a_j, H_I\right] \nonumber\\
  & = & -\hbar\frac{d\phi_j^a(t)}{dt}\tilde{a}_j + J_j^{ab}(t)\tilde{b}_j + J_{j-1}^{ba}(t)\tilde{b}_{j-1}
\end{eqnarray}
with $J_j^{ab}(t)=\hbar\Omega_j^{ab}(t)/2$ and $J_j^{ba}(t)=\hbar\Omega_j^{ab}(t)/2$. Similarly, for operator $\tilde{b}_j$, we have
\begin{equation}
 i\hbar\frac{d \tilde{b}_j}{dt} = -\hbar\frac{d\phi_j^b(t)}{dt}\tilde{b}_j + J_j^{ab}(t)\tilde{a}_j + J_{j+1}^{ba}(t)\tilde{a}_{j+1}~~.
\end{equation}
Since the equations of motion for the operators $\{\tilde{a}_j, \tilde{b}_j\}$ should also follow $i\hbar\frac{d \tilde{a}_j}{dt} = [\tilde{a}_j, \tilde{H}_I]$, $i\hbar\frac{d \tilde{b}_j}{dt} = [\tilde{b}_j, \tilde{H}_I]$, now we can write the effective Hamiltonian as
\begin{equation}
\begin{split}
 \tilde{H}_I = & \sum_j\left(-\hbar\frac{d\phi_j^a(t)}{dt}\tilde{a}^\dagger_j\tilde{a}^{\phantom\dagger}_j -\hbar\frac{d\phi_j^b(t)}{dt}\tilde{b}^\dagger_j\tilde{b}^{\phantom\dagger}_j\right) \\ & + \sum_j\left[ \left(J_j^{ab}(t)\tilde{a}^\dagger_j \tilde{b}^{\phantom\dagger}_j + J_j^{ba}(t)\tilde{b}^\dagger_j\tilde{a}^{\phantom\dagger}_{j+1}\right) + {\rm h.c.}\right] \ ,
 \end{split}
\end{equation}
from which we can write the potential difference between neighboring sites as 
\begin{equation}
    \Delta (t) = \hbar\left( \frac{d\phi_j^a(t)}{dt} - \frac{d\phi_j^b(t)}{dt}\right) = \hbar\frac{d\phi_j^{ab}(t)}{dt} \ .
\end{equation}
Then, we have the following form of the phases for each MW tone, 
\begin{equation}
    \phi_j^{ab}(t) = \phi_j^{ab}(t=0) + \frac{1}{\hbar}\int_0^t \Delta(t')dt' \ .
\end{equation}
The initial phases $\phi_j^{ab}(t=0)$ do not affect the derivation of the effective potential difference between neighboring sites, and we simply set them to zero in our experiment.

\section{SPAM error}\label{appendixB}
As discussed in \cite{Chen2024}, the primary data we measure for the state population dynamics has a lower contrast than the renormalized data presented in the main text. Two main effects reduce the contrast
of the raw population dynamics data. First, the data typically features an average upper “ceiling” value $P_u$, which
stems from the STIRAP inefficiency and loss during release-and-recapture. There is also a lower baseline of the
measurements, having an average value $P_l$, that we believe stems from the decay (and subsequent recapture) of the
short-lived Rydberg states, which results in the non-depumped Rydberg states having some probability to appear
bright to subsequent fluorescence detection. These infidelities limit the contrast of state population dynamics. 

\begin{figure}[t]
	\includegraphics[width=0.5 \textwidth]{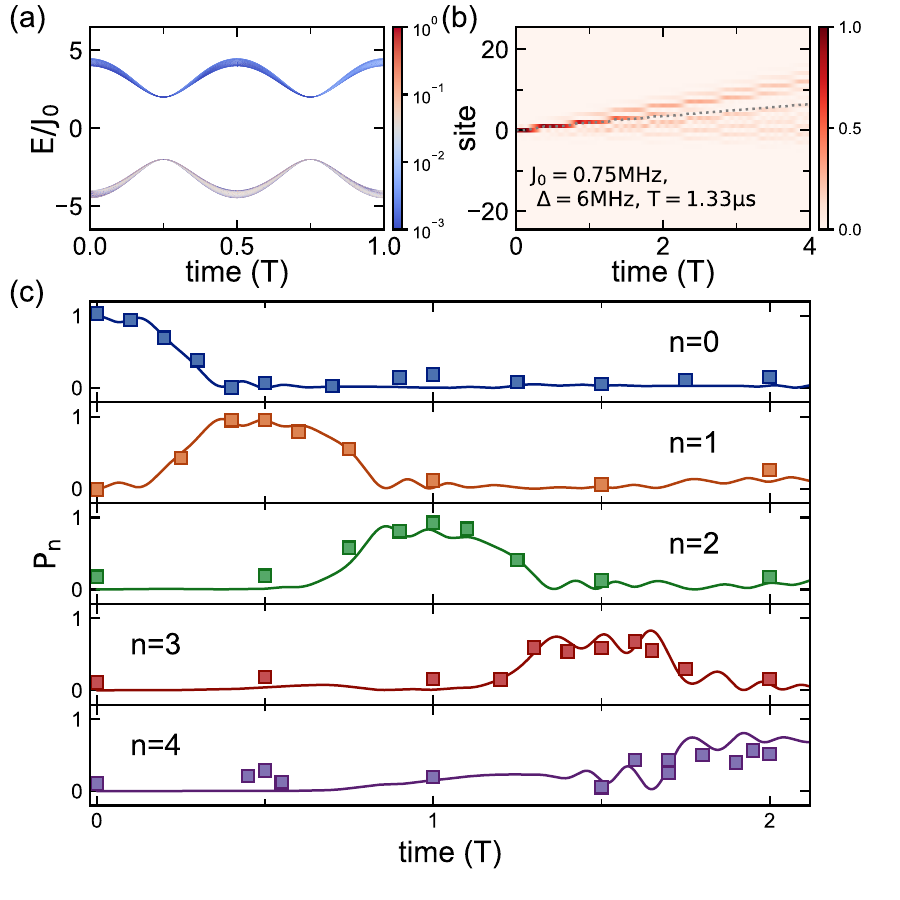}
	\caption{\textbf{Single particle pump with $\Delta = 8J_0 = h \times 6$~MHz}
 \textbf{(a)}~A plot of the instantaneous/adiabatic energy spectrum throughout a pumping period for the Rice-Mele model with larger imbalance-to-hopping ratio. 
 The color indicates the overlap between time-evolved state vector and the instantaneous eigenstates. 
 As compared to Fig.~\pref{FIG:fig4}{a}, the initial projection onto a single band is better for the larger $\Delta/J_0$. For the two-atom case, this projection for the initial state $|00\rangle$ gives an improved overlap with the lower band of $\sum|\langle \Psi_{l}|00\rangle|^2 = 0.97$. \textbf{(b)}~In a 50-site simulation, the population still spreads out at longer times due to the finite width of the individual bands, but the center of mass position $\lambda = 6.4$ over 4 pumping periods gives the pumping efficiency $\lambda/(2t/T) = 0.8$ that exceeds that of the non-interacting pump considered in the main text.
 \textbf{(c)}~Experimental measurements under this pump in the 5-site synthetic lattice, showing agreement with theory (solid lines) and a more regular pumping along the synthetic lattice, stemming in part from the improved band projection and in part from the more ``control freak''~\cite{Asbóth2016} nature of this pump trajectory.}

\label{FIG:figS1}
\end{figure}

For the averaged population dynamics in non-interacting singles $(P_i = \langle c^{\dagger}_i c^{\phantom\dagger}_i \rangle)$ and the interacting pairs $[P_i = \frac{1}{2}(\langle c^{\dagger}_{i,A} c^{\phantom\dagger}_{i,A} \otimes I_B \rangle +  \langle I_A \otimes c^{\dagger}_{i,B} c^{\phantom\dagger}_{i,B}\rangle)]$, we renormalize the measured $P^{\text{bare}}_i$ to $P_i = (P^{\text{bare}}_i - P_i)/(P_u - P_i)$ with $P_u = 0.90(1)$ and $P_l = 0.30(3)$ for the $^{39}$K set up and $P_u = 0.90(1)$ and $P_l = 0.10(2)$ for the $^{87}$Rb set up. To note, when performing this normalization we systematically do not account for the statistical
variations of the renormalization factors, which will lead to additional (and unaccounted for) uncertainties on the
values of the renormalized population data. We also note that,
as the value of $P_l$ depends on the decay of the Rydberg states, we operate our experiments with a fixed duration of time between the initial Rydberg state excitation and the subsequent de-excitation (a duration of 5~$\mu$s), to help ensure a fixed $P_l$ value for all the data, independent of evolution time under the synthetic lattice Hamiltonian. Our detection (based on imaging ground state atoms) always happens at a fixed time point in the sequence.

\begin{figure}[b]
	\includegraphics[width= 0.5 \textwidth]{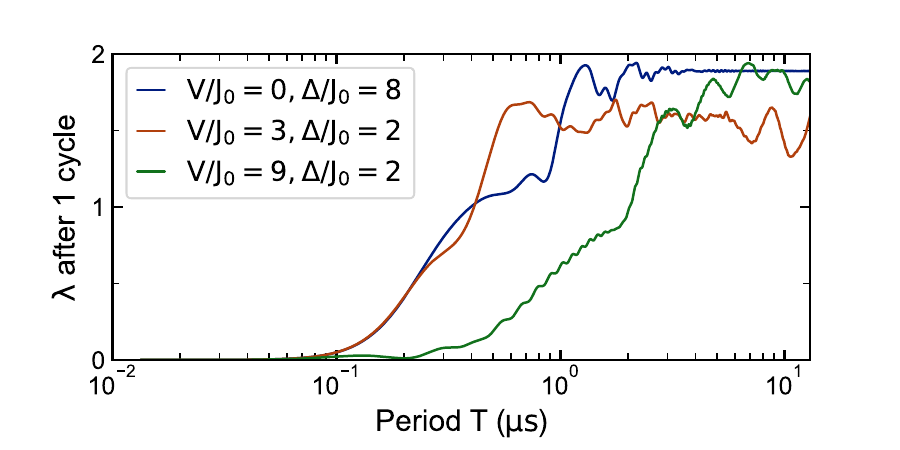}
	\caption{\textbf{Center of mass (COM) position for two atoms after one pumping cycle as a function of the pumping period.} Simulations are performed with a 102-site system size for three conditions as labeled in the plot. We note that the COM takes a value of $\lambda = 2$ for an ideal pump with perfect band projection after one pumping period. All three curves approach the ideal pumping behavior with larger pumping period $T$. Due to the finite system size, edge interference effects contaminate the pumping efficiency for longer $T$ values.}
\label{FIG:figS2}
\end{figure}

\section{Single particle pump with better initial state projection}\label{appendixC}
In the main text, operating with moderate ratios of the imbalance-to-hopping parameters $\Delta/J_0 \in \{2,2.8\}$, the initialization of our single-particle pump from site-localized atoms suffers the non-ideality of an imperfect projection onto a single-band of the Rice-Mele model. That is, for these $\Delta/J_0$ ratios and for our initial pump phase, the single-band Wannier states are not purely site-localized. To achieve a more ideal initial projection, such that only the lower band is populated, different approaches can be taken. In the simulations of the ``ideal'' pumping curves in Fig.~\ref{FIG:fig1}, we also considered the case of a different initial phase of the pump cycle to ensure more fully dimerized conditions. Experimentally, we also considered pumping in the case of larger potential imbalance values (larger $\Delta/J_0$ ratios), and observing more idealized single-particle pumping as expected. Figure~\ref{FIG:figS1} provides an analysis of the effective band structure of the dimensionally extended Rice-Mele model under this pump [panel (a)], the expected synthetic lattice pumping dynamics for an enlarged system size [panel (b)], and the experimental site population dynamics of single atoms under a few periods of pumping along with a comparison to simulations for our 5-site synthetic lattice [panel (c)].

\section{Numerical investigation of pump adiabaticity}\label{appendixD}
We performed simulations with the experimentally relevant pump trajectories/parameters to further explore the adiabaticity of the pumps for single particles and interacting pairs. Here we explore the case of the pumping trajectory as used for pairs of $^{39}$K atoms in the main text. The numerically simulated results are shown in Fig.~\ref{FIG:figS2} for the cases of $V/J_0 = 0$ (single particles) and for moderate ($V/J_0 = 3$) and large ($V/J_0=9$) interactions.
We plot the center-of-mass position $\lambda$ after one pump period (and additionally starting with atoms at the central site of a larger lattice to try to mitigate boundary reflection effects for slow pumping rates) as a function of the pumping period (with values of $J_0$ the same as in the main text for the cases of $^{39}$K pumping), where the ideal pump would provide a center-of-mass position of $\lambda = 2$.
One sees that as interaction increases, especially for $V/J_0=9$, much longer pumping periods are required due to the reduced bandwidth in the effective pair-hopping Rice-Mele model.

We note that, in performing simulations of three all-to-all interacting atoms with $V \gg J_0$, it was practically impossible to reach the regime of adiabatic pumping on accessible timescales (limited by system size, namely effects of small population reaching the system boundary).

This points to a qualitative distinction between two-atom pumps and many-atom pumps, namely a transition to self-trapping and soliton-like behavior. While pairs of atoms can support effective pair-basis dynamics with modest bandwidths, the corresponding bandwidth of many-atom pumps will be significantly reduced due to an exponential suppression of multi-atom hopping. Specifically, for $N$ all-to-all-connected particles, the corresponding bandwidth scales as $\sim$$J_0 (J_0/V)^{(N-1)}$, making it practically impossible (for large $N$) to maintain adiabaticity in the $V/J_0 \gg 1$ limit. This effective freeze-out of multi-particle dynamics and the emergence of effectively self-trapped states signals one crossover from the few- to many-body regimes of this problem.

\section{Simulations for pumping with $V\gg J_0$}\label{appendixE}
\begin{figure}[t]
	\includegraphics[width= 0.5 \textwidth]{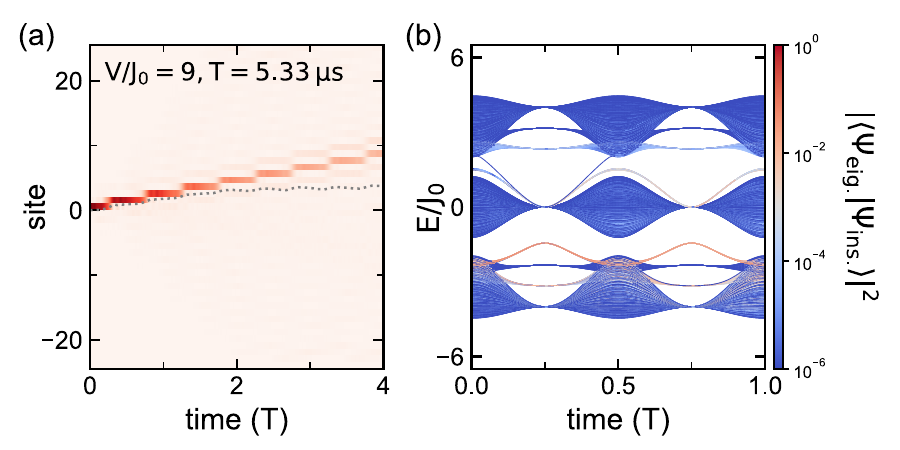}
	\caption{\textbf{Simulation with large interactions ($V/J_0 = 9$) and slow modulation ($T = 5.33$~$\mu$s).} (a)~Pumping dynamics in the large interaction limit is revived with a slower pump rate. To note, the center-of-mass position (dashed line) turns over at long times due to an interference effect with population reflecting from the system boundaries. (b)~Instantaneous energy structure showing that less population transfers to the the upper sub-band after the minimum gap point at time $t = T/4$, as compared to the case of a faster pumping rate in Fig.~\ref{FIG:fig4}. Color indicates the overlap of the time-evolved state vector with the instantaneous eigenstates.
    }
\label{FIG:figS3}
\end{figure}

As suggested by Fig.~\ref{FIG:fig4} and the effective pair-state Rice-Mele model, the pumping trajectory of pair states remains topological, as it is in the single-particle case. However, for large $V$, smaller pumping rates are required to maintain adiabaticity (as also shown in Fig.~\ref{FIG:figS2}) and to observe pumping in spite of this decreasing bandwidth (relating to the decreasing rate of pair-hopping for increasing $V/J_0$). Fig.~\ref{FIG:figS3} shows the simulated population dynamics and the two-particle energy structure for a pair of atoms in the $V\gg J_0$ limit, in the case of a pumping period that is much longer than that considered in the main text. Here, the pumping in space [panel (a)] is improved for larger pumping periods and there is less transfer between the energy sub-bands associated with the effective pair basis states [panel (b)] as compared to Fig.~\pref{FIG:fig4}{c}.

\begin{figure}[b]
	\includegraphics[width= 0.5 \textwidth]{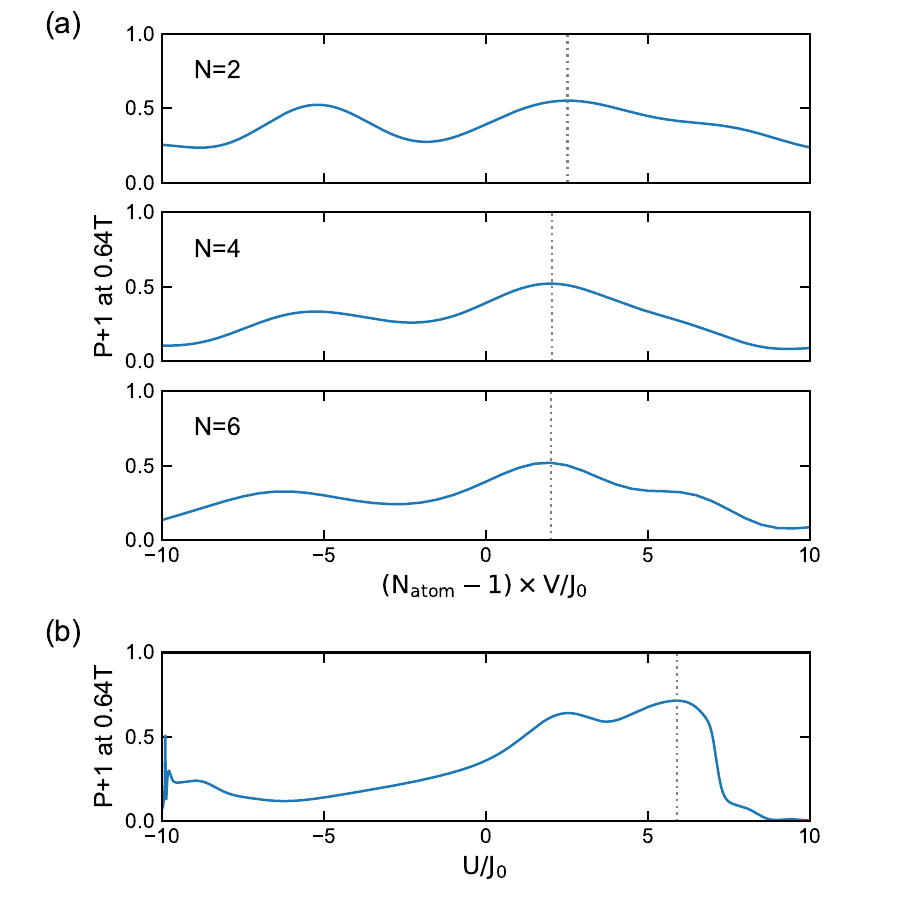}
	\caption{\textbf{Numerical simulations comparing all-to-all connected quantum pumping dynamics with analogous nonlinear classical wave simulations.} The curves show the population pumped to one site above the initial site at a half pump period (with the same pump parameters as in Fig.~\ref{FIG:fig7 ?}), indicative of the tendency for atoms to be self-bound while pumping along the internal dimension. In the simulation, we assume center site initialization in a 9-site synthetic lattice with exchange interactions that are uniform between neighboring states along the internal dimension. Two types of interactions are explored: \textbf{(a)}~atoms experience all-to-all interactions in real space with the number of atoms labeled on the plots; \textbf{(b)} atoms experience a collective real-space interaction in the classical limit, with $U$ describing a nonlinear interaction with the population at neighboring synthetic sites. The dotted vertical lines indicate the interaction-to-hopping ratios that yield the peak pumping probability.}

\label{FIG:figS4}
\end{figure}

\section{Comparison between infinite-range quantum pumps and their semi-classical equivalent}\label{appendixF}

In the main text, in considering the pump dynamics of few- and many-atom arrays, we discuss apparent similarities with the behavior of nonlinear photonic pumps~\cite{Jurgensen2021}. We provide some arguments for this similarity: the long-ranged interactions of dipolar Rydberg arrays in real space~\cite{Tromb-RMP} may make them similar to the collective Kerr nonlinearity, and there is a local (when reduced to a two-mode system) equivalence of the Kerr nonlinearity and the flip-flop exchange interactions. Here, we provide a more detailed comparison between the pumping behavior of all-to-all connected, globally driven quantum pumps and pumps of a classical wave system subject to an off-site nonlinearity that is akin to the flip-flop exchange interaction if one traces over the spatial degree of freedom. While in experiment we are restricted to studying all-to-all interactions for just three atoms in a triangle, Fig.~\pref{FIG:figS4}{a} considers up to six particles with uniform and full connections. Here, we consider the same pumping parameters as used for the cases of $^{87}$Rb pumping in Fig.~\ref{FIG:fig7 ?}, and we plot the probability for atoms to be pumped forward by one site as a function of the interaction strength. Here, we plot versus the normalized interaction strength $V/J_0$ that is additionally scaled by the effective connectivity $N_\textrm{atom}-1$ of the various all-to-all clusters. Roughly speaking, independent of $N$, the peak of the pumping probability occurs when the collective interaction energy per atom is a few times the hopping energy $J_0$.

Figure~\pref{FIG:figS4}{b} then compares to nonlinear wave equation simulations that are identical with the quantum simulations at the linear or single-particle level (with $H^\textrm{s.p.}$ the single-particle portion of the Hamiltonian), but which incorporate interactions between population at the ``sites'' $j$ along the synthetic dimension of the form 
$i\hbar\dot\psi_j = \sum_i H^\textrm{s.p.}_{i,j}\psi_j + U (|\psi_{j-1}|^2 + |\psi_{j+1}|^2) \psi_j$. The interaction-dependence of the pumping of population along the synthetic dimension is qualitatively quite similar to that of the quantum cases - optimal pumping for positive and moderate interaction-to-hopping ratios and with some smaller enhancement at negative interactions. Even some of the more specific details of the individual site population dynamics (not depicted) show qualitative similarities to the few-atom simulations of dipolar Rydberg atoms subject to moderate interactions (with qualitative differences for positive and negative interactions).

\bibliographystyle{apsrev4-1}
%

\clearpage

\renewcommand{\thesection}{\Alph{section}}
\renewcommand{\thefigure}{S\arabic{figure}}
\renewcommand{\thetable}{S\Roman{table}}
\setcounter{figure}{0}
\renewcommand{\theequation}{S\arabic{equation}}
\renewcommand{\thepage}{S\arabic{page}}
\setcounter{equation}{0}
\setcounter{page}{1}

\end{document}